\documentclass[twocolumn]{aastex63}
\usepackage{ amsmath}

\newcommand{\be}{\begin{equation}}
\newcommand{\ee}{\end{equation}}

\shorttitle{Dust Echoes from Luminous Fast Blue Optical Transients}
\shortauthors{B.D.~Metzger, D.~Perley}
\graphicspath{{./}{figures/}}

\begin{document}

\title{Dust Echoes from Luminous Fast Blue Optical Transients}


\author[0000-0002-4670-7509]{Brian D.~Metzger}
\affil{Department of Physics and Columbia Astrophysics Laboratory, Columbia University, New York, NY 10027, USA}
\affil{Center for Computational Astrophysics, Flatiron Institute, 162 5th Ave, New York, NY 10010, USA}

\author[0000-0001-8472-1996]{Daniel Perley}
\affil{Astrophysics Research Institute, Liverpool John Moores University, IC2, Liverpool Science Park, 146 Brownlow Hill, Liverpool L3 5RF, UK}

\begin{abstract}
Luminous fast blue optical transients (LFBOTs) such as AT2018cow form a rare class of engine-powered explosions of uncertain origin.  A hallmark feature of these events is radio/millimeter synchrotron emission powered by the interaction of fast $\gtrsim 0.1 c$ ejecta and dense circumstellar material (CSM) extending to large radii $\gtrsim 10^{16}$ cm surrounding the progenitor.  Assuming this CSM to be an outflow from the progenitor, we show that dust grains up to $\sim 1\mu m$ in size can form in the outflow in the years before the explosion.  This dusty CSM would attenuate the transient's ultraviolet (UV) emission prior to peak light, before being destroyed by the rising luminosity, reddening the pre-maximum colors (consistent with the pre-maximum red-to-blue color evolution of the LFBOT candidate MUSSES2020J).  Re-radiation by the dust before being destroyed generates an near-infrared (NIR) ``echo'' of luminosity $\sim 10^{41}-10^{42}$ erg s$^{-1}$ lasting weeks, which is detectable over the transient's rapidly fading blue continuum.  We show that this dust echo is compatible with the previously unexplained NIR excess observed in AT2018cow.  The gradual decay of the early NIR light curve can result from CSM which is concentrated in wide-angle outflow or torus, consistent with the highly aspherical geometry of AT2018cow's ejecta.  Pre-maximum optical/UV and NIR follow-up of LFBOTs provide an new probe of their CSM environments and place additional constraints on their progenitors.

\end{abstract}

\section{Introduction}

``Fast Blue Optical Transients" (FBOTs; \citealt{Drout+14,Arcavi+16,Pursiainen+18,Ho+21b}) are UV-bright supernova(SN)-like explosions of debated origin characterized by UV/optical rise times of only days and peak luminosities up to $\gtrsim 10^{44}$ erg s$^{-1}$.  This paper focuses on the most luminous FBOTs (hereafter ``LFBOTs''), which are extremely rare ($\lesssim 0.1-1\%$ of the core-collapse SN rate; e.g., \citealt{Coppejans+20,Ho+21b}) and likely possess distinct progenitors from the majority of FBOTs (whose properties may form a continuum with previously-identified SN classes; \citealt{Ho+21b}).  

The prototypical LFBOT is AT2018cow, which exhibited multi-wavelength emission spanning radio to gamma-rays \citep{Prentice+18,RiveraSandoval+18,Kuin+19,Perley+19,Margutti+19,Ho+19,Nayana&Chandra21}.  AT2018cow's UV/optical light curve rose over just a few days to a peak luminosity $L \approx 4\times 10^{44}$ erg s$^{-1}$ before declining rapidly thereafter.  This emission was accompanied by variable non-thermal X-rays present from the first observations days after the explosion (e.g., \citealt{RiveraSandoval+18,Kuin+19}).  The X-rays likely originate from a compact object central engine \citep{Perley+19,Margutti+19,Pasham+21} whose spectra (namely time-dependent Fe and Compton-hump emission features) indicate is embedded within the expanding ejecta shell \citep{Margutti+19}.  The optical wavelength spectra were initially featureless over the first couple weeks after the explosion, indicating large expansion velocities $v \gtrsim 0.1 c$ and high temperatures $\approx 3\times 10^{4}$ K.  Later spectra revealed the emergence of H and He emission features at significantly lower velocities $v \approx 3000-4000$ km s$^{-1}$ with no evidence for ejecta cooling (e.g., \citealt{Perley+19,Margutti+19,Xiang+21}).  

Narrow He emission lines $\lesssim 300$ km s$^{-1}$ were also detected from AT2018cow, pointing to the presence of dense H-depleted circumstellar material (CSM) ahead of the ejecta (e.g., \citealt{Fox&Smith19,Dessart+21}).  Indeed, one of the hallmark features of LFBOTs is the presence of luminous radio and millimeter synchrotron emission \citep{Ho+19,Margutti+19,Nayana&Chandra21} arising due to shock interaction between the fastest ejecta (velocity $v \gtrsim 0.1$ c) and CSM of density $n \gtrsim 10^{6}-10^{7}$ cm$^{-3}$ on radial scales $\sim 10^{16}$ cm surrounding the progenitor.  

Over the past few years, additional LFBOTs have been discovered which exhibit qualitatively similar multi-wavelength properties to AT2018cow, including CSS161010 \citep{Coppejans+20}, AT2018lug (``Koala"; \citealt{Ho+20}), AT2020xnd \citep{Bright+21,Perley+21,Ho+21}, and AT2020mrf  \citep{Yao+21}, albeit with a wide range of X-ray luminosities and fastest ejecta speeds implied by their radio/mm emissions.  An LFBOT-like optical/UV transient, MUSSES2020J, was recently discovered which reached an even higher peak luminosity $\sim 10^{46}$ erg s$^{-1}$ with the final light curve rise by a factor of $\gtrsim 100$ captured over just 5 days \citep{Jiang+22}, though the identification of this event as an LFBOT is not certain because of the lack of multi-wavelength data.  To summarize the physical picture that has emerged from detailed modeling (e.g., \citealt{Margutti+19}): AT2018cow and other LFBOTs are engine-powered explosions which generate highly aspherical ejecta spanning a wide range of velocities, propagating into a dense, radially-extended pre-existing CSM.  The densest portion of the ejecta, and potentially also of the CSM, is concentrated in an equatorial torus, presumably orthogonal to the rotational or orbital angular momentum axis of the progenitor system.

A large number of progenitor models for LFBOTs have been proposed, including: the SN explosion of a rapidly-rotating massive star with a low total ejecta mass giving birth to a central engine, such as a millisecond magnetar or black hole (\citealt{Prentice+18,Perley+19,Margutti+19,Fang+19,Gottlieb+22}); an initially ``failed'' SN which nevertheless produces an accreting black hole and mass ejection \citep{Quataert+19,Perley+19,Margutti+19,Antoni&Quataert22}; the tidal disruption of a star by an intermediate-mass black hole \citep{Kuin+19,Perley+19} or a stellar-mass black hole in a dense stellar environment \citep{Kremer+21}; pulsational pair instability SNe (e.g., \citealt{Leung+20}); and the merger of a helium core with a black hole or neutron star following a common envelope event \citep{Soker+19,Schroder+20,Uno&Maeda20,Soker22,Metzger22}.

Here we point out that, regardless of its origin, the CSM surrounding LFBOTs is likely to be sufficiently cool and dense prior to the explosion to facilitate the formation and growth of dust.  This dusty CSM would furthermore be opaque to the transient's early UV emission before peak light, until the dust is destroyed by the rising luminosity.  This has two observable consequences: (a) the pre-maxima UV/optical spectra of LFBOTs will be substantially redder than the bluer colors observed near and after peak light; (b) the UV luminosity absorbed during the early rising phase prior to dust destruction, will be re-emitted by the heated dust at near-infrared (NIR) wavelengths, giving rise to a ``dust echo'' lasting weeks.  As we will show, such an echo appears compatible the heretofore mysterious NIR excess observed from AT2018cow by \citet{Perley+19}.  The discovery of similar and related dust signatures in future LFBOTs offers a new probe of their large-scale progenitor environments complementary to those provided by radio/mm observations.

\begin{figure}
    \centering
    \includegraphics[width=0.5\textwidth]{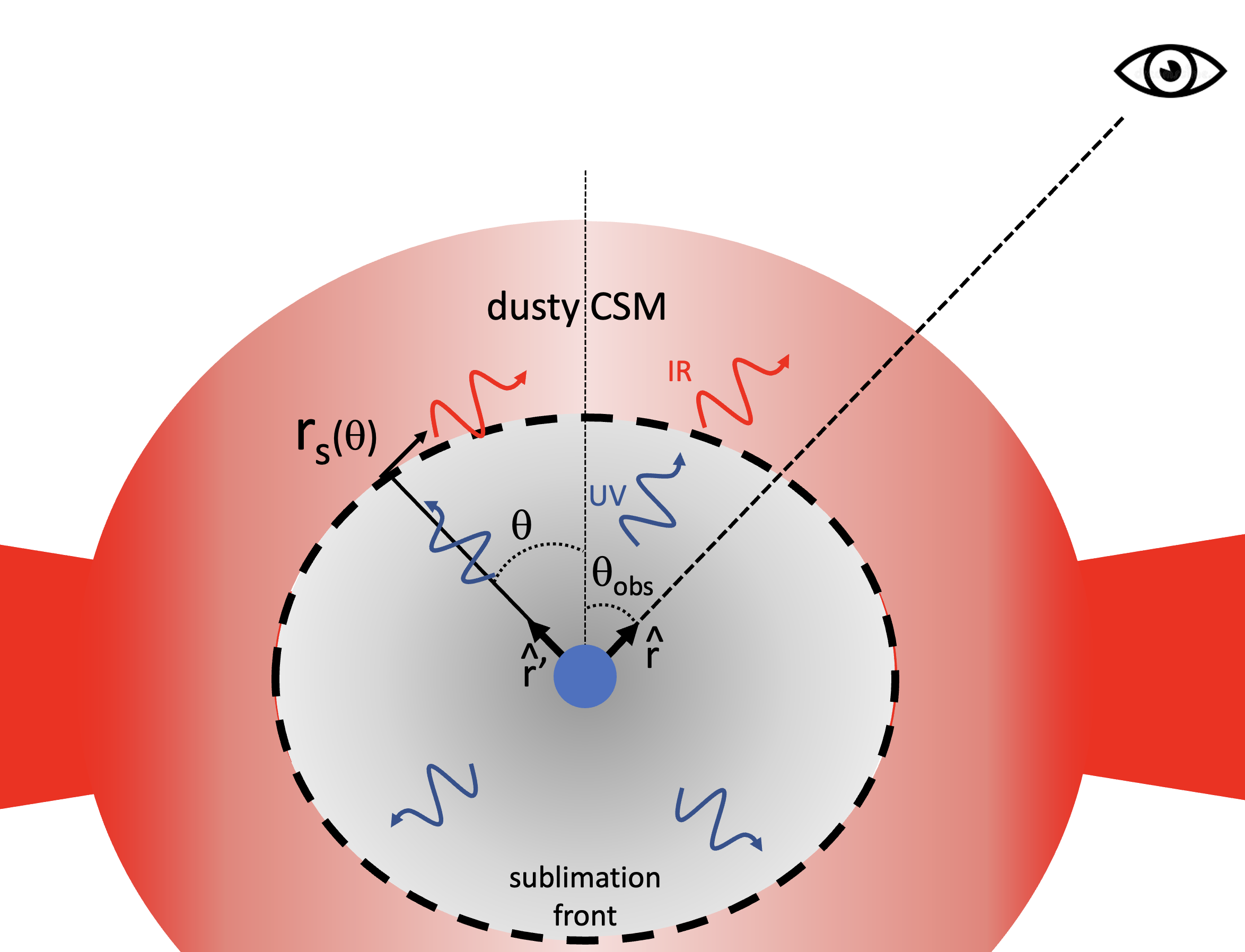}
    \caption{Schematic illustration of the geometry of dust destruction and light echos in LFBOTs.  Modeling of AT2018cow and other LFBOTs reveals dense CSM extending to radii $\gtrsim 10^{16}$ cm surrounding the progenitor system, potentially concentrated in an equatorial torus along the same symmetry axis as inferred from the highly-aspherical ejecta.  The CSM is sufficiently dense to form and grow dust grains outside the condensation radius $r_{\rm c} \gtrsim 10^{14}-10^{15}$ cm (Eq.~\ref{eq:rc}).  At early times following the explosion, the dusty CSM is opaque to UV/optical emission (optical depth $\tau_{\rm UV} \gg 1$).  As the UV luminosity rises to maximum over a few days, the dust is heated and destroyed out to increasingly large radii, eventually, reducing $\tau_{\rm UV} < 1$.  Before being destroyed, the heated dust re-radiates the absorbed UV energy, most of which is emitted at the radius $\sim r_{\rm thin} \gtrsim 10^{16}$ cm (Eq.~\ref{eq:rthin}) where $\tau_{\rm UV} \sim 1$ (but $\tau_{\rm IR} \ll 1$).  An observer viewing the system from an angle $\theta_{\rm obs}$ with respect to the CSM polar axis receives the IR emission from radius $r_{\rm s}$ and angle $\theta$ with a time delay $\Delta t_{\rm IR} \simeq (r_{\rm thin}(\theta)/c)\left(1-\hat{r}'\cdot \hat{r}\right)$, set by the light-crossing time, where $r_{\rm thin}(\theta) \propto n^{1/2}$ grows with the polar angle due to the increasing density $n(\theta)$ (see Eq.~\ref{eq:ntheta}).  Reprocessed emission may not reach the observer from the bottom hemisphere ($\theta > \pi/2$) if the dense equatorial disk is opaque to the NIR emission ($\tau_{\rm IR} > 1$).}
    \label{fig:cartoon}
\end{figure}

\section{Dust in LFBOTs}

We begin in Sec.~\ref{sec:dust_formation} by summarizing what is known about the CSM properties surrounding LFBOTs and use this information to estimate the dust formation properties in this environment.  In Sec.~\ref{sec:destruction} we describe the conditions for dust destruction during the LFBOT rise phase, before estimating the properties of the NIR echo in Sec.~\ref{sec:echo}.  The geometry of the system is illustrated in Fig.~\ref{fig:cartoon}.

\subsection{Dust Formation in the Progenitor CSM}
\label{sec:dust_formation}

The synchrotron radio/millimeter emission from LFBOTs starting weeks after the explosion is powered by shock interaction of a fast outflow of velocity $v \approx 0.1-0.5$ c with dense CSM (e.g., \citealt{Ho+19,Margutti+19,Coppejans+20,Nayana&Chandra21,Bright+21,Margalit+22}).  Application of standard synchrotron shock models to two well-studied events, AT2018cow and AT2020xnd, found similar CSM density profiles \citep{Bright+21}, which on radial scales $r \sim 10^{16}$ cm are roughly approximated as a power-law of the form (\citealt{Bright+21}; their Fig.~4):
\be
n = n_7 \times 10^{7}{\rm cm^{-3}} \left(\frac{r}{10^{16}\,{\rm cm}}\right)^{-3},
\label{eq:n}
\ee
where the precise normalization $n_{7} \sim 0.1-1$ depends on the shock microphysics (e.g., the fraction of the shock energy placed into magnetic fields, $\epsilon_{\rm B}$).  Although such a steep density profile is not compatible with that of a steady wind ($n = \dot{M}_{\rm w}/4\pi r^{2} m_p v_{\rm w} \propto r^{-2}$), its normalization corresponds to a mass-loss rate $\dot{M}_{\rm w} \sim 10^{-3}-10^{-2}M_{\odot}$ yr$^{-1}$ for an assumed wind velocity $v_{\rm w} = 300$ km s$^{-1}$ \citep{Fox&Smith19}.  Modeling the optical/X-ray emission from LFBOTs (e.g., \citealt{Margutti+19}) indicates the ejecta from the explosion, and potentially also the CSM, is not distributed uniformly surrounding the progenitor, but is instead concentrated in a equatorial outflow or thick torus with a higher density $n(\theta)$ at larger angles $\theta$ (Fig.~\ref{fig:cartoon}) from the polar axis (along which the explosion ejecta is fastest and the radio/mm emission predominantly originates; see Sec.~\ref{sec:echo} for further discussion).

Stellar objects undergoing high rates of mass-loss in other contexts, such as giant outbursts from luminous blue variables, are observed to generate copious amounts of dust (e.g., \citealt{Davidson&Humphreys97,Kochanek11}).  We now estimate the size of the dust grains that could grow in the outflows of LFBOT progenitors of density profile \eqref{eq:n}.  Motivated by the oxygen-rich composition of massive-star outflows, we focus on silicate grains.  However, qualitatively similar results would follow for carbonaceous grains, and when possible we interpret our analytic estimates for dust properties (e.g., sublimation temperatures) spanning both possibilities.  

Once the temperature in the progenitor outflow decreases from its initially hot state close to the progenitor to values $T < T_{\rm c}$ at radii $\sim r_{\rm c}$, solid condensation occurs and grain nucleation can begin (e.g., \citealt{Cherchneff&Dwek09,Nozawa+14}), where $T_{\rm c} \approx 1200\,{\rm K}(1500-2000\,{\rm K})$ for silicate(carbonaceous) grains, respectively.

Assuming that grain growth is dominated by the rate of accumulating monomers, the timescale for a spherical dust grain to grow to radii $a = a_{\mu m}\mu m$ is given by (e.g., \citealt{Kwok75}; see \citealt{Kochanek11}, their Eq.~17),
\begin{eqnarray}
t_{\rm grow} &\equiv& \frac{a}{\dot{a}} \approx \frac{4 \rho_{\rm b} a}{n(r_{\rm c}) m_p X_{\rm d}v_{\rm c}(T_{\rm c})} \nonumber \\
&\approx& 0.26\,{\rm yr} \,\,\frac{a_{\mu m}}{n_{7}X_{\rm d,-1}T_{\rm c,1200}^{1/2}}\left(\frac{r_{\rm c}}{10^{15}{\rm cm}}\right)^{3}, 
\label{eq:tgrow}
\end{eqnarray}
where in the second line we take $\rho_{\rm b} \approx 3.8$ g cm$^{-3}$ as the bulk density of the silicate dust grain and $X_{\rm d} = 0.1 X_{\rm d,-1}$ is the mass fraction of dust-generating condensible material, normalized to a higher value than for solar metallicity given the hydrogen-depleted nature of the CSM surrounding LFBOTs (e.g., \citealt{Perley+19}).  The effective collisional velocity $v_{\rm c}$ is assumed to be the thermal velocity,
\be
v_{\rm c} = \left(\frac{8 kT}{\pi m_a}\right)^{1/2} \simeq 1.1\times 10^{5}{\rm cm\,s^{-1}}T_{1200}^{1/2},
\ee
where $T_{1200} = T/(1200$ K) and we take $m_a = 20 m_p$ for the monomer building blocks of silicate grains \citep{Kochanek11}.  

The radius $r_{\rm c}$ at which $T = T_{\rm c}$ entering Eq.~\eqref{eq:tgrow} depends on the radiation environments of the LFBOT progenitors, which are uncertain.  One clue may come from the luminous UV source ($L > 10^{40}$ erg s$^{-1}$; $T_{\rm eff} > 4\times 10^{4}$ K) with $H\alpha$ emission features detected by the {\it Hubble Space Telescope} from the location of AT2018cow $\sim 2-3$ yrs after the explosion \citep{Sun+22}; the photometric stability of this source over two epochs disfavors an origin associated with the transient itself, instead pointing to an extremely massive $\gtrsim 100M_{\odot}$ stellar binary or cluster companion to AT2018cow's progenitor.  Assuming the progenitor was a massive star of comparable luminosity $L_{\star} \sim 10^{40}$ erg s$^{-1}$, the radiation temperature $T_{\rm rad} = (u_{\rm rad}/a)^{1/4}$ in the CSM with energy density $u_{\rm rad}$ at radius $r$ in the outflow can be estimated as\footnote{One way to understand the factor $(\tau_{\rm IR}+1)$ which enters $u_{\rm rad}$ is because radially-constant energy flux $F_{\rm rad} = u_{\rm rad}v_{\rm diff}$ will be established between the location of interest (optical depth $\tau_{\rm IR} > 1$) and the photosphere ($\tau_{\rm IR} \approx 1$) at which $u_{\rm rad}(\tau_{\rm IR} = 1) = L_{\star}/4\pi c r^{2}$ and $v_{\rm diff} \approx c/\tau_{\rm IR}$ is the radial diffusion velocity of the IR photons.}
\begin{eqnarray}
T_{\rm rad} &\simeq& \left(\frac{L_{\star}(\tau_{\rm IR}+1)}{4\pi \sigma r^{2}}\right)^{1/4} \nonumber \\
&\approx& 1941\,{\rm K}\left(\frac{L_{\star}(\tau_{\rm IR}+1)}{10^{40}{\rm erg\,s^{-1}}}\right)^{1/4}\left(\frac{r}{10^{15}{\rm cm}}\right)^{-1/2},
\end{eqnarray}
where $\tau_{\rm IR} \sim 1$ is the radial optical depth through the CSM to the re-radiated NIR emission of temperature $\sim T_{\rm rad}$.  Assuming the gas and radiation are in equilibrium ($T \simeq T_{\rm rad}$), this results in a condensation radius (where $T \simeq T_{\rm c}$) of
\be
r_{\rm c} \approx 2.6\times 10^{15}\,{\rm cm}\,\frac{(\tau_{\rm IR}+1)^{1/2}}{T_{\rm c,1200}^{2}} \left(\frac{L_{\star}}{10^{40}{\rm erg\,s^{-1}}}\right)^{1/2}.
\label{eq:rc}
\ee

The size to which dust grains can grow is determined by a comparison between the grain growth timescale and the outflow (or inflow) timescale from the condensation radius $t_{\rm exp} \sim r_{\rm c}/v_{\rm w}$ (e.g., \citealt{Kochanek11}).  Evaluating Eq.~\eqref{eq:tgrow} at $r = r_{\rm c}$,
\be
\frac{t_{\rm grow}}{t_{\rm exp}} \sim  1.7\frac{a_{\mu m}(\tau_{\rm IR}+1)}{n_{7}X_{\rm d,-1}T_{\rm c,1200}^{9/2}}\left(\frac{v_{\rm w}}{300\,{\rm km\,s^{-1}}}\right)\left(\frac{L_{\star}}{10^{40}{\rm erg\,s^{-1}}}\right),
\ee
where $v_{\rm w}$ is normalized to a value $300$ km s$^{-1}$ close to the narrow spectral-line features in AT2018cow (e.g., \citealt{Fox&Smith19}).  Comparing $t_{\rm grow}$ with $t_{\rm exp}$, we see that grain growth to sizes $a > 0.1-10\mu m$ is possible ($t_{\rm grow}/t_{\rm exp} < 1$), for reasonable ranges in the values of $\{n_{7}, v_{\rm w}$, $L_{\star}$, $T_{\rm c}$, $\tau_{\rm IR}\}$.   

The dusty outflow at radii $\gtrsim r_{\rm c}$ will obscure the progenitor at optical/UV wavelengths.  The absorption opacity in the geometric optics limit (wavelengths $\lambda \ll 2\pi a$), is approximately given by
\be
\kappa_{\rm UV} = \frac{3}{4}\frac{X_{\rm d}}{\rho_{\rm b}a} \approx 2.0\times 10^{2}\,{\rm cm^{2}\,g^{-1}}\frac{X_{\rm d,-1}}{a_{\mu m}}.
\label{eq:kappaUV}
\ee
The optical depth of the CSM dust to UV radiation external to a given radius $r$ is
\begin{eqnarray}
\tau_{\rm UV}(>r) \simeq \int_{r}^{\infty}m_p n \kappa_{\rm UV}dr \nonumber \\
\approx 16.5 \frac{n_7 X_{\rm d,-1}}{a_{\mu m}}\left(\frac{r}{10^{16}{\rm cm}}\right)^{-2},
\label{eq:tauUV}
\end{eqnarray}
where we have used Eqs.~\eqref{eq:n}, \eqref{eq:kappaUV}.  Absent dust destruction, the UV emission from LFBOTs should be obscured ($\tau_{\rm UV} \gtrsim 1$) by the dusty CSM on the same radial scales $\sim 10^{16}$ cm probed by the transient's radio/mm emission, contrary to observations taken near and after the optical/UV peak showing blue colors from a high-temperature $\gtrsim 10^{4}$ K continuum.  

\subsection{Dust Destruction by the Rising UV Transient}
\label{sec:destruction}

As the UV luminosity $L_{\rm UV}$ of the transient rises to its peak, this radiation will heat and sublimate the dust to increasingly large radii, reducing the optical depth (Eq.~\ref{eq:tauUV}).  To estimate the dust sublimation radius, we follow similar considerations in the context of other explosive transients: SNe (e.g., \citealt{Bode&Evans80,Dwek83,Maeda+15}), AGN (e.g., \citealt{Barvainis87}), gamma-ray bursts (e.g., \citealt{Waxman&Draine00,Waxman+22}) and tidal disruption events \citep{vanVelzen+16}.  We compute the temperature of the dust by equating the rate a grain of radius $a$ at radius $r$ absorbs heat,
\be
L_{\rm abs} = Q_{\rm UV}\frac{L_{\rm UV}\pi a^{2}}{4\pi r^{2}},
\ee
where $Q_{\rm UV} \sim 1$ is the absorption efficiency at UV wavelengths where the spectral energy distribution (SED) of LFBOT peaks at early times ($T_{\rm eff} \approx 3\times 10^{4}$ K; e.g., \citealt{Perley+19}), to the rate at which dust radiates 
\be
L_{\rm dust} = 4\pi a^{2} \langle Q_{\rm IR}\rangle_{\rm T} \sigma T_{\rm d}^{4},
\ee
where $Q_{\rm IR}$ is the emission efficiency.  For silicate grains across a range of sizes $a \sim 0.01-1 \mu m$ the Planck-averaged emissivity at temperatures $T_{\rm d} \sim 10^{3}$ K is approximately given by (e.g., \citealt{Draine&Lee84}, their Fig.~11)
\be
\langle Q_{\rm IR}\rangle_{\rm T} \approx 0.3 a_{\mu m}\left(\frac{T_{\rm d}}{2300{\rm K}}\right)
\ee
Setting $L_{\rm abs} = L_{\rm dust}$ gives the dust temperature,
\begin{eqnarray}
T_{\rm d} &\simeq& \left(\frac{L_{\rm UV}}{16\pi \sigma \langle Q_{\rm IR}\rangle_{\rm T}  r^{2}}\right)^{1/4} \nonumber \\
&\approx& 3066\,{\rm K}\,L_{43}^{1/5}a_{\mu m}^{-1/5}\left(\frac{r}{10^{16}{\rm cm}}\right)^{-2/5},
\end{eqnarray}
where $L_{43} \equiv L_{\rm UV}/(10^{43}\,\rm erg\,s^{-1})$.  Following Eq.~12 of \citet{Waxman&Draine00}, we estimate that for UV irradiation lasting from hours to days during the transient rise (see below), dust will be destroyed by sublimation above the temperature $T_{\rm d} = T_{\rm s} \approx 1700$ K, at the radius
\be
r_{\rm s} \approx 4.4\times 10^{16}{\rm cm}\,\, L_{43}^{1/2}a_{\mu m}^{-1/2}T_{\rm s,1700}^{-5/2}.
\label{eq:rs}
\ee
A somewhat higher sublimation temperature $T_{\rm s} \approx 2000$ K would characterize carbonaceous grains.  The UV optical depth through the CSM when the luminosity reaches $L_{\rm UV}$ is thus given by (Eq.~\ref{eq:tauUV}) 
\be
\tau_{\rm UV}(>r_{\rm s}) =  0.84 \, n_{7}L_{43}^{-1}X_{\rm d,-1}T_{\rm s,1700}^{5}.
\ee
As the sublimation radius grows, the value of $\tau_{\rm UV}$ is reduced to $\lesssim 1$ once the luminosity increases to a value
\be
L_{\rm thin} \simeq 8.4\times 10^{42}{\rm erg\,s^{-1}}n_{7}X_{\rm d,-1}T_{\rm s,1700}^{5},
\label{eq:Lthin}
\ee
which is notably independent of the grain size.  This occurs at a critical radius
\be
r_{\rm thin} = r_{\rm s}(L_{\rm thin}) \approx 4.1\times 10^{16}{\rm cm}\, n_{7}^{1/2}X_{\rm d,-1}^{1/2}a_{\mu m}^{-1/2}.
\label{eq:rthin}
\ee

\subsection{Re-radiated IR Emission}
\label{sec:echo}

Grains heated by the transient just prior to their destruction will re-radiate the absorbed energy at the temperature of the dust grains.  In particular, most of the energy absorbed when $\tau_{\rm UV} \gtrsim 1$ ($L < L_{\rm thin}$) will be emitted as quasi-blackbody emission of temperature $T \approx T_{\rm sub} \approx 1700-2000$ K in the NIR.  The total UV radiation energy absorbed by the dust on radial scales $\sim r_{\rm thin}$ (Eq.~\ref{eq:rthin}) is estimated as that absorbed when $\tau_{\rm UV} \sim 1$,
\be
E_{\rm IR} \approx L_{\rm thin} \Delta t_{\rm L_{\rm thin}}(\Delta \Omega/4\pi),
\label{eq:EIR}
\ee
where $\Delta t_{\rm L_{\rm thin}}$ is the time the transient spends during its rise around the luminosity $\sim L_{\rm thin}$ and $\Delta \Omega$ is the total solid angle subtended by the CSM (we assume the transient's UV luminosity is isotropic across $\Delta \Omega$).  

Until recently, no LFBOT UV/optical light curves were available in the literature which start significantly before maximum light to enable a measurement of the rise.  We accordingly adopt a pre-maximum rise $L_{\rm UV} \propto t^{2}$ motivated by early SN observations (e.g., \citealt{Olling+15}), i.e. 
\be
L_{\rm UV} = L_{\rm pk}\left(\frac{t}{t_{\rm pk}}\right)^{2},\,\,\, t < t_{\rm pk}
\ee
where $t$ is measured from the onset of the explosion and $L_{\rm pk}$ and $t_{\rm pk}$ are the peak luminosity and peak/rise-time of the transient, respectively.  This rise law is roughly consistent with that measured for the extremely luminous LFBOT-like transient MUSSES2020J (\citealt{Jiang+22}), though we caution that early dust attenuation could act to steepen the observed rise compared to the intrinsic one (we return to possible observational evidence for such early attenuation in Sec.~\ref{sec:conclusions}).

The time spent around a given luminosity is then
\be
\Delta t_{\rm L} \equiv \frac{L_{\rm UV}}{dL_{\rm UV}/dt} = \frac{t_{\rm pk}}{2}\left(\frac{L_{\rm UV}}{L_{\rm pk}}\right)^{1/2}.
\ee
Evaluating this for $L_{\rm UV} = L_{\rm thin}$ (Eq.~\ref{eq:Lthin}), Eq.~\eqref{eq:EIR} becomes:
\begin{eqnarray}
&& E_{\rm IR} \sim \frac{L_{\rm thin}^{3/2}}{L_{\rm pk}^{1/2}}\frac{t_{\rm pk}}{2}\frac{\Delta \Omega}{4\pi} \nonumber \\
&\approx& 5.3\times 10^{47}\,{\rm erg}\, n_{7}^{3/2}X_{\rm d,-1}^{3/2}t_{\rm pk,5}L_{\rm pk,44}^{-1/2}(\Delta \Omega/4\pi),
\end{eqnarray}
where we have normalized $L_{\rm pk} = L_{\rm pk,44}\times 10^{44}$ erg s$^{-1}$ and $t_{\rm pk} = 5 t_{\rm pk,5}$ days to typical values for the peak luminosity and decay time\footnote{Though not a perfect point of comparison, the u-band light curves of Type Ib/c SNe exhibit similar rise and decay timescales to within a factor $\lesssim 2$ (see \citealt{Taddia+15}; their Table 3).}, respectively, of LFBOT light curves (e.g., \citealt{Prentice+18}).  

Because the dust grains are distant from the source of emission, there is a significant time delay before the NIR radiation reaches a distant observer (Fig.~\ref{fig:cartoon}).  If the direction from the center of the explosion to the observer is the polar axis of a system of spherical coordinates, the delay is given by
\be
\Delta t_{\rm IR} = (R/c)(1-\cos \tilde{\theta}),
\ee
where $R$ is the radial coordinate to a particular location on the dust shell and $\tilde{\theta}$ is its polar angle.\footnote{The angle $\tilde{\theta}$ with respect to the observer line of sight is not to be confused with the polar angle $\theta$ relative to the CSM symmetry axis (Fig.~\ref{fig:cartoon})}  Insofar as $E_{\rm IR}$ is emitted from a shell of radius $R = r_{\rm thin}$ (Eq.~\ref{eq:rthin}), this gives a maximal delay ($\tilde{\theta} = \pi, \Delta \Omega = 4\pi$)
\be
\Delta t_{\rm IR,max} = \frac{2 r_{\rm thin}}{c} \approx 31.6\,{\rm d}\,n_{7}^{1/2}X_{\rm d,-1}^{1/2}a_{\mu m}^{-1/2}.
\label{eq:tIR}
\ee
The response function of a spherical shell (or a portion of a spherical shell axisymmetric about the line of sight to the observer) which re-radiates light simultaneously is a square-wave, resulting in a flat light curve of luminosity
\begin{eqnarray}
&& L_{\rm IR} \approx  \frac{E_{\rm IR}}{\Delta t_{\rm IR}} =
 \frac{E_{\rm IR}(\Delta \Omega = 4\pi)}{\Delta t_{\rm IR, max}} \nonumber \\
&\approx& 1.8\times 10^{41}{\rm erg\,s^{-1}}\,n_7 X_{\rm d,-1}a_{\mu m}^{1/2}L_{\rm pk,44}^{-1/2} t_{\rm pk,5}
\label{eq:LIR}.
\end{eqnarray}
The plateau of luminosity $L_{\rm IR}$ lasts a duration $\Delta t_{\rm IR} < \tau_{\rm max}$, depending on the angular extent of the dust shell.  

More generally the light curve shape will not be strictly flat, if either (a) the CSM is not distributed in a spherically symmetric manner around the explosion, i.e. the CSM density varies as a function of the polar angle with respect to some symmetry axis; or (b) reprocessed light reaches the observer from only a portion of the total 4$\pi$ solid angle, for instance if the CSM in the equatorial plane (Fig.~\ref{fig:cartoon}) blocks reprocessed NIR light from the opposite hemisphere.  Both of these effects play a role in our model for AT2018cow's NIR echo in the next section.

\section{Application to AT2018cow}

\begin{figure}
    \centering
    \includegraphics[width=0.48\textwidth]{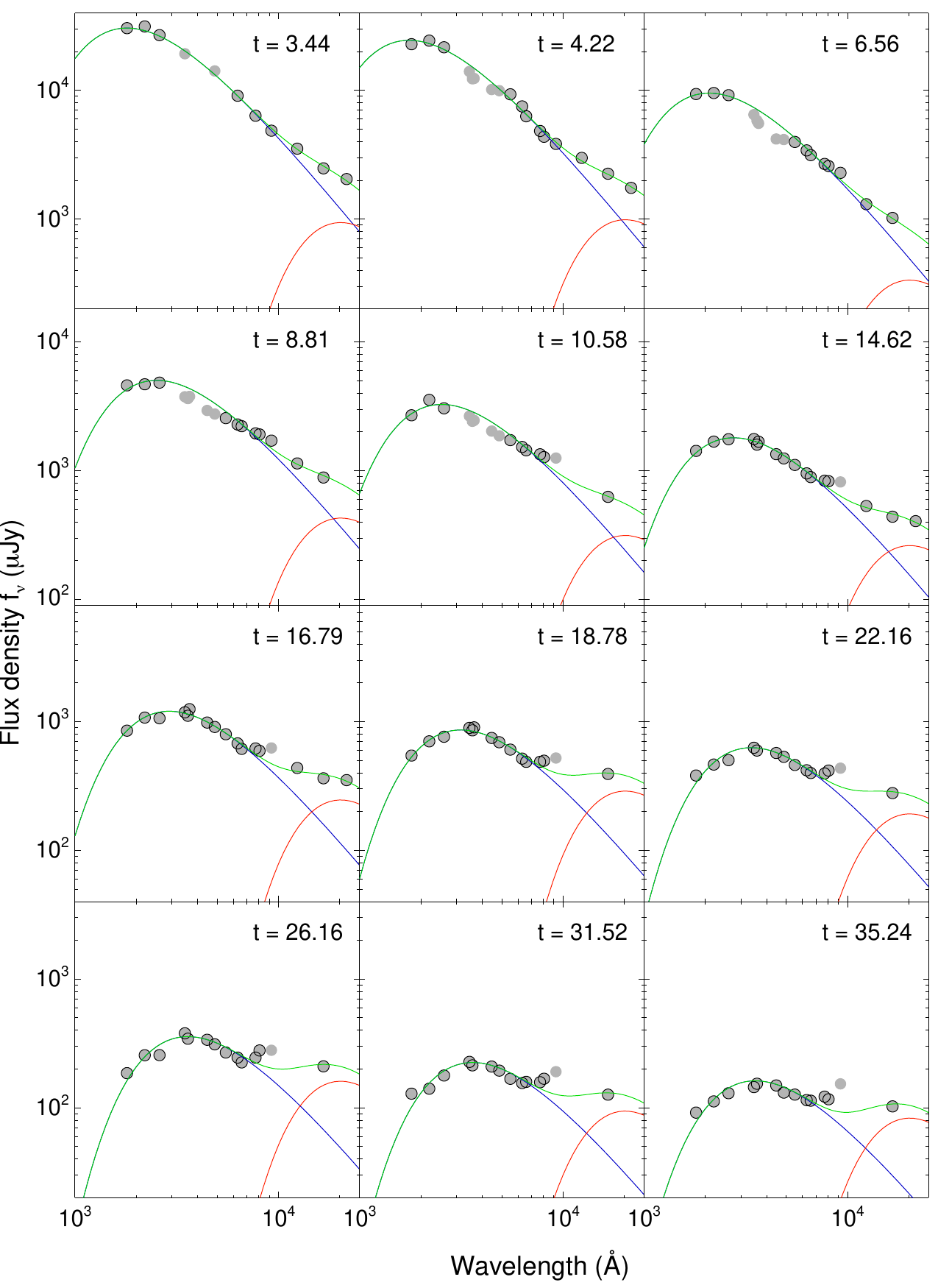}
    \caption{Two-component fits to the time-dependent SEDs of AT\,2018cow for a selection of 12 epochs.  The blue component shows the primary blackbody, the solid red component shows the dust graybody, and the green curve is the sum of these.  Points with no outline were excluded from fitting.}
    \label{fig:sedfits}
\end{figure}

\begin{figure}
    \centering
    \includegraphics[width=0.5\textwidth]{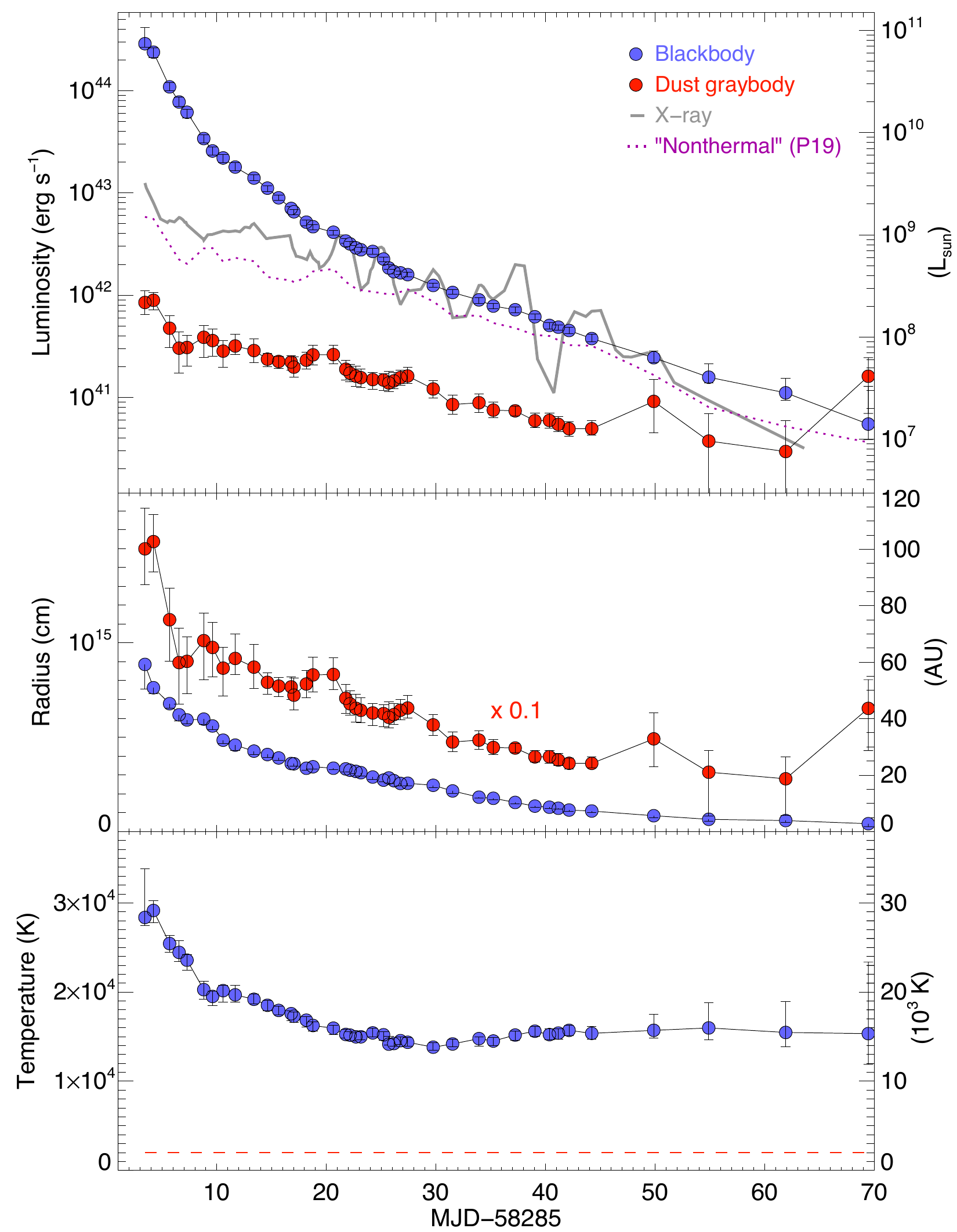}
    \caption{Time-evolution of model parameters for the blackbody+graybody fit.  Blue points show the properties of transient photosphere; red points show the emission properties of the putative dust shell (the effective radius has been scaled down by a factor of 10).  The temperature of the dust component is fixed to 2000\,K.  The top panel also shows the X-ray luminosity and the luminosity of the power-law component that was used to fit the NIR excess in \cite{Perley+19}.}
    \label{fig:sedparameters}
\end{figure}

Although most of its luminosity is radiated in the UV and optical bands, \citet{Perley+19} identified in AT2018cow a source of NIR emission in excess of the best-fit blackbody fit to the UV/optical continuum (see their Figs. 7, 8; we reproduce these data here in Figs.~\ref{fig:sedfits}, \ref{fig:sedparameters}).  This excess shows up in the spectrum most prominently around $\lambda \approx 2 \mu m$ and possesses a much flatter light curve than the rapidly-fading UV/optical continuum, remaining present for at least 44 days.

\citet{Perley+19} consider a non-thermal origin for this NIR excess, arguing it is part of the same synchrotron emission spectrum responsible for the millimeter band flux.  However, \citet{Margutti+19} showed that an extrapolation of the model that best fits the radio observations at $\nu \ll 100$ GHz severely underpredicts the NIR flux (the NIR band is likely above the synchrotron cooling break).  It is thus tempting to identify AT2018cow's NIR excess as a dust echo similar to those explored here.  

\citet{Perley+19} considered the possibility of dust emission for the NIR excess in AT2018cow, but disfavored this explanation on the grounds that: (a) dust emission lacks an physical origin (now provided in this paper); and (b) fitting the UVOIR data with two blackbodies results in best-fit temperature for the NIR component closer to 3000 K, hotter than would be expected for dust.  However, the spectrum of the dust echo emission will be that of a {\it modified} blackbody $B'_{\nu} \approx B_{\nu}Q_{\rm IR} \simeq B_{\nu}\nu^{q}$ of temperature $T \simeq T_{\rm s} \approx 1700-2000$ K, where $q \approx 1-1.5$ at NIR wavelengths for micron-sized grains (\citealt{Draine&Lee84}; their Fig.~5) with a spectral peak at $h\nu_{\rm pk} \approx (3.9-4.5)kT_{\rm s}$ ($\lambda \approx 2 \mu m$).  We note that if one tries to fit the peak of the spectrum assuming pure blackbody emission (for which $h\nu_{\rm pk} \approx 2.8kT$) this would lead to an overestimate of the temperature by a factor of $\approx 1.4-1.6$, transforming the $\approx$ 3000 K best-fit blackbody spectrum inferred by \citet{Perley+19} to one with $\approx 2000$ K, closer to the emission temperature of soon-to-be sublimated dust. 

In order to explore this issue, we repeated the SED fits using a modified blackbody equation of the type above in place of the nonthermal power-law originally employed in \citet{Perley+19}.  The temperature was fixed to 2000 K and the blackbody curve was attenuated by an attenuation factor of the form $Q(\lambda) = ((\frac{\lambda}{a})^{-sq}) + \frac{\lambda}{a})^{-1/s}$ \citep{Beuermann+99}.  The break wavelength $a$ was set to 1 $\mu$m, the break sharpness parameter to $s=2.0$, and the long-wavelength power-law index $q=1$.  The modified emission curve was then numerically integrated at each epoch to calculate the luminosity.  The resulting fits are shown in Figure \ref{fig:sedfits}, and the time-evolution of the parameters is presented in Figure \ref{fig:sedparameters}.

\begin{figure}
    \centering
    \includegraphics[width=0.5\textwidth]{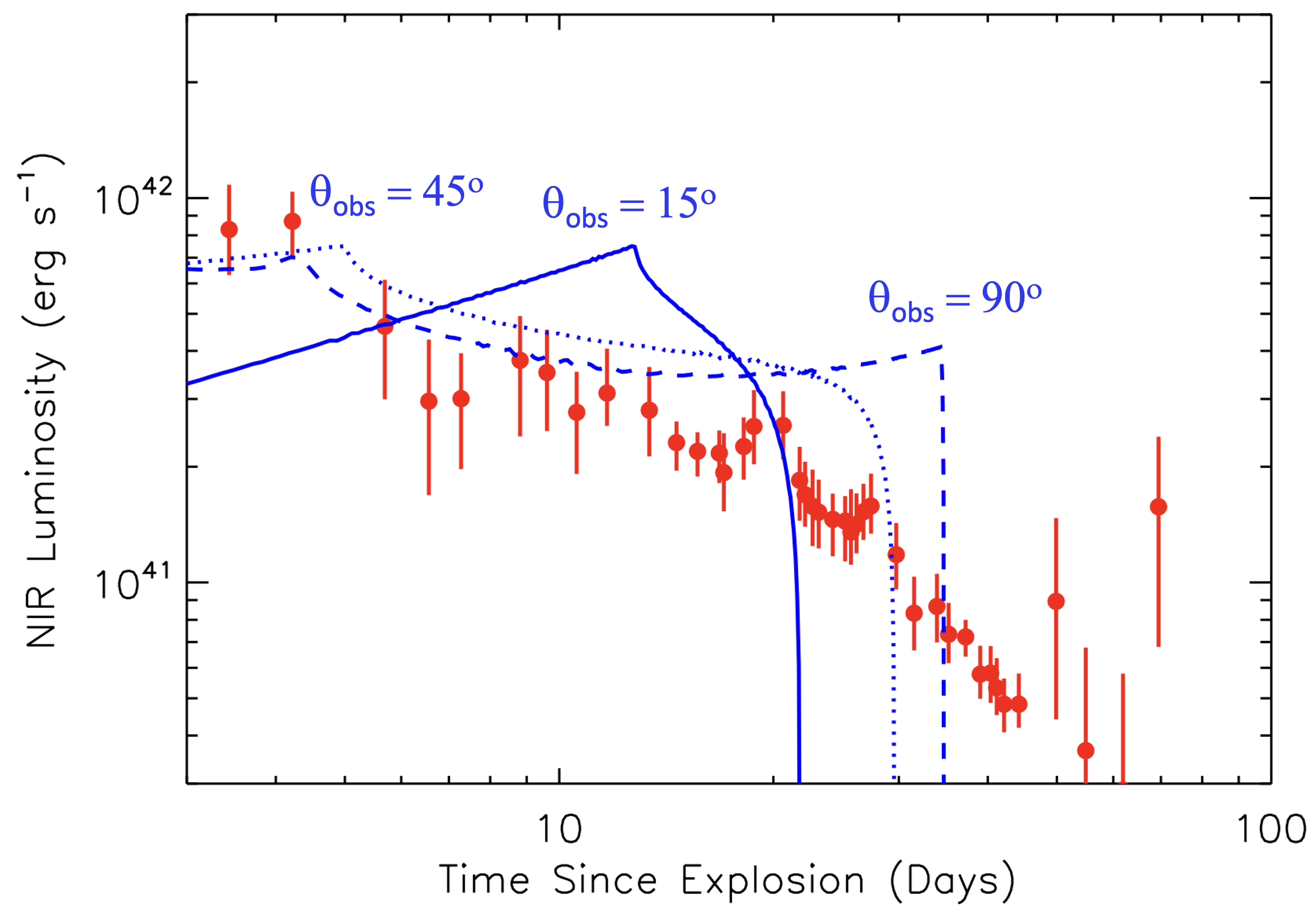}
    \caption{Blue lines show models for the NIR dust echo luminosity for different observing angles $\theta_{\rm obs}$ with respect to the symmetry axis of the CSM density distribution (Eq.~\ref{eq:ntheta}; see Fig.~\ref{fig:cartoon} for the assumed geometry).  The normalization of the luminosity requires CSM properties: $n_0(X_{\rm d}/0.1) \sim 3\times 10^{7}$ cm$^{-3}$, $a \sim 1 \mu m$ (see Eq.~\ref{fig:luminosity}).  The model only includes dust emission from radii $\sim r_{\rm thin}$ and the hemisphere closest to the observer ($\theta > \pi/2$).  For comparison, we show the excess NIR emission component from AT2018cow as red points (Fig.~\ref{fig:sedparameters}).}
    \label{fig:luminosity}
\end{figure}

This revised model shows a qualitatively good match to the data, with a median residual of less than 0.05 mag across all epochs, comparable to the expected systematic error.  The goodness-of-fit is inferior to that of the original power-law model in \cite{Perley+19} ($\chi^2$ of 225 and 145, respectively, on 370 degrees of freedom), although the difference originates primarily from an apparent flux excess in the $i$ and $z$ bands at late times, which may be due to an unidentified spectral feature.  Both models do an excellent job of fitting the early-time data before this feature appears, notably including the epochs at 14.6 and 16.8 days where $JHK$ photometry available.

\subsection{NIR Light Curve Modeling}

Eqs.~\eqref{eq:tIR}, \eqref{eq:LIR} show that both the characteristic timescale and luminosity of the measured NIR emission ($L_{\rm IR} \sim 10^{41}-10^{42}$ erg s$^{-1}$; $\Delta t_{\rm IR} \approx 30-60$ d; Fig.~\ref{fig:luminosity}) can be explained for dust grain sizes $a \gtrsim \mu$m and an assumed CSM density $(n_0/10^{7}{\rm cm^{-3}})(X_{\rm d}/0.1) \sim 3$, moderately higher than is found by radio modeling of AT2018cow and other LFBOTs (\citealt{Ho+19,Margutti+19}).  As we now discuss, however, this high density is consistent given the angular distribution of the CSM inferred from the NIR light curve shape.  

As has already been mentioned, neither the ejecta velocity nor the CSM it interacts with are likely to be spherically symmetric, as significant pole-to-equator gradients in both are needed to explain the multi-wavelength data from AT2018cow \citep{Margutti+19,Fox&Smith19} and are predicted by several LFBOT progenitor models as a result of stellar rotation or binarity (e.g., \citealt{Quataert+19,Soker+19,Schroder+20,Uno&Maeda20,Metzger22}).  Thus, those regions of the CSM probed most sensitively by the radio/mm emission may not match those which dominate the bulk of the reprocessed IR emission.  In particular, the ejecta-CSM shock-powered synchrotron emission is more sensitive to the ejecta speed than CSM density and hence will preferentially arise from higher latitudes (small $\theta$ in Fig.~\ref{fig:cartoon}) along which the fast jet-like polar outflow from the central engine is directed.  By contrast, the reprocessed IR luminosity $L_{\rm IR} \propto n$ will arise preferentially from closer to the equatorial plane ($\theta \approx \pi/2$) where the CSM density is greatest.

A similar angular dependence for the CSM density could contribute to the shape of AT2018's NIR light curve, particularly its gradual decay over the first 30 days after the explosion (versus the strictly flat light curve prediction for a spherical dust shell).  
To illustrate this, we have calculated the dust echo light curve for an assumed CSM density profile of the form,
\be
n(\theta) = n_0 \sin^{2}\theta,
\label{eq:ntheta}
\ee
which peaks in the equatorial plane ($\theta = \pi/2$).  This profile results in an angle-dependent dust emission radius $r_{\rm thin}(\theta) = r_{\rm thin,0}\sin\theta \propto n^{1/2}$ (Eq.~\ref{eq:rthin}).  We calculate the NIR emission received by an observer at a given polar angle $\theta_{\rm obs}$ by summing the contributions from each solid angle $d\Omega = d\phi d(\sin \theta)$ subtended by the central explosion weighting it by the reprocessed energy $E_{\rm IR}(\theta) \propto r_{\rm thin}^{3}$ (Eq.~\ref{eq:EIR}) along that direction, accounting for the light-travel arrival delay $\Delta t = (r_{\rm thin}/c)(1- \hat{r}\cdot \hat{r}')$ (see Fig.~\ref{fig:cartoon} for the adopted geometry).  The assumed grain size $a_{\mu m} \sim 1\mu m$ and mid-plane density $n(\theta = \pi/2) = n_0 \sim 3\times 10^{7}$ cm$^{-3}$ determine through $r_{\rm thin,0} \approx 5\times 10^{16}$ cm and the overall normalization of $E_{\rm IR}$ the characteristic duration and peak luminosity of the emission, which still roughly follow the analytic estimates in Eqs.~\eqref{eq:tIR}, \eqref{eq:LIR} (provided one uses the peak CSM density, $n = n_0$).  We only consider emission from the CSM at angles $0 < \theta < \pi/2$, assuming the dense equatorial CSM will attenuate reprocessed emission from the opposing hemisphere ($\tau_{\rm IR} > 1$).

The brown lines in Figure \ref{fig:luminosity} show our calculation for different observer viewing angles, $\{\theta_{\rm obs} = \pi/12, \pi/4, \pi/2\}$, in comparison to the NIR light curve of AT2018cow.  For typical observing angles (e.g., $\theta_{\rm obs} \approx \pi/4$), one predicts a roughly flat brief initial light curve phase (emission from dust closest to the observer's line of sight, where the CSM is locally homogeneous for small angles around $\theta \approx \theta_{\rm obs}$) followed by gradual decay over longer timescales as emission is received from larger and larger angles approaching the opposite side of the hemisphere.  

The lack of detailed agreement between the model and the observed IR light curve is not unexpected given our ad hoc CSM density profile (Eqs.~\ref{eq:n}, \ref{eq:ntheta}) and other simplifications of the model (we do not account for angle-dependent dust grain properties and radiative transfer effects, for example).  The late-time NIR emission at $t \gtrsim$ 30 days is also not captured by the toy model, though such emission can arise from dust at radii $> r_{\rm thin}$ (where $\tau_{\rm UV}<1$ so less energy is reprocessed with a longer delay time; e.g., \citealt{Maeda+15}) or from the opposite hemisphere ($\pi/2 < \theta < \pi$) that nevertheless reaches the observer despite absorption by the equatorial CSM.

In our interpretation for the NIR excess from AT2018cow as a dust echo, the broadly similar shape of the X-ray and NIR light curves in Fig.~\ref{fig:sedfits} would be coincidental.  It is tempting to consider whether the X-ray emission itself could be scattered emission from an early short-lived extremely X-ray-luminous phase.  However, such an interpretation is challenged by the unique time-dependent spectral evolution and variability \citep{Margutti+19} and QPE emission \citep{Pasham+21}.  Another point of tension regarding the reprocessing scenario are the apparent short-timescale variability/wiggles in the NIR light curve, such as the ``bump'' at $t \approx 20$ days.  A clumpy CSM with a large density enhancement along the observer's line of site at radii $\gg r_{\rm thin}$, might be needed to generate such features.

\section{Summary}
\label{sec:conclusions}

We conclude that$-$whatever their nature$-$the progenitors of LFBOTs just before exploding, may find themselves enshrouded in dust formed in their own dense CSM, the presence of which is made apparent by the luminous shock-powered radio/mm emission following the explosion.  In principle, such dust-shrouded progenitors could be observable as IR sources with luminosities up to $\sim 10^{40}$ erg s$^{-1}$ (if the putative companion star to AT2018cow detected by \citealt{Sun+22} is representative of its progenitor), qualitatively similar to the IR-luminous progenitors of the SN2008S-like explosions (e.g., \citealt{Prieto+08,Thompson+09}).  Given the short-lived phase of the CSM mass-loss ($\lesssim 10$ yr; see below) and extremely low rate of LFBOTs, such ``ultraluminous IR sources'' would be extremely rare and difficult to discover in the local universe.  We estimate that the planned High Latitude Wide Area Survey on the Nancy Grace Roman Space Telescope (\citealt{Wang+22}) could in principle detect such a progenitor at the $\approx$60 Mpc distance of AT2018cow if it emits a significant fraction of its luminosity at $\lambda \approx 1-2\mu$m.

Though most of the dust is ultimately destroyed by the transient's UV light, the few percent of the radiated energy which absorbed during the rise before this point is re-emitted as a NIR echo of luminosity $L_{\rm IR} \sim 10^{41}-10^{42}$ erg s$^{-1}$ lasting weeks to months.  Given the long wavelength of this emission (a modified blackbody of effective temperature $\lesssim T_{\rm s} \approx 1700-2000$ K), this signal is visible over the transient's rapidly-fading optical/UV continuum.  We encourage future NIR follow-up of LFBOTs as a probe of the CSM properties on radial scales $\sim 10^{16}$ cm complementary to that provided by radio/mm observations.  

The high luminosity and early-time gradual-decay of AT2018cow's IR light curve arise naturally if the CSM is preferentially concentrated in a thick equatorial outflow or torus with a peak density $n(r \sim 10^{16}$ cm$) > 10^{8}$ cm$^{-3}$ for $X_{\rm d} < 0.1$ greater than that at higher latitudes probed by the radio/mm shock emission by a factor $\gtrsim 10$ (Eq.~\ref{eq:n}).  If the CSM is an outflow, its corresponding mass-loss rate for $v_{\rm w} \sim 300$ km s$^{-1}$ must be large $\dot{M}_{\rm w} \sim 0.1M_{\odot}$ yr$^{-1}$ and it must contain a total mass $\gtrsim 0.1-1M_{\odot}$  Such a large quantity of mass-loss years prior to the explosion provides a strong constraint on any LFBOT progenitor model (see \citealt{Metzger22} for a discussion).

Dust signatures including excess IR emission is observed in many core-collapse SNe (e.g., \citealt{Bode&Evans80,Dwek83,Fox+13}), particularly in a handful of the Type Ibn SNe subclass (e.g., \citealt{Smith+08,Mattila+08,Gan+21}) spectrally most similar to AT2018cow.  However, the dust emission in these events does not generally become prominent for several weeks or longer after the explosion and the emitting dust may be freshly synthesized by the shock rather than being exclusively pre-existing CSM (e.g., \citealt{Mattila+08,Smith+08}).  By contrast, the IR emission component in AT2018cow is already present even days after peak light, when the equilibrium temperature behind even the fastest portions of the ejecta is still too high $\gtrsim 10^{4}$ K for dust nucleation.

Though challenging giving their rarity and rapid evolution, we encourage efforts to discover LFBOTs and obtain multi-band photometry if not spectra during their pre-maximum rise phase.  The optical/UV spectrum should be substantially reddened by dust attenuation up to luminosities $L_{\rm thin} \gtrsim 10^{43}-10^{44}$ erg s$^{-1}$ (Eq.~\ref{eq:Lthin}) capable of destroying the dust.  

A possible test of this prediction came with the discovery of the extremely luminous LFBOT candidate MUSSES2020J \citep{Jiang+22} four days prior to the optical maximum.  Contrary to the naive expectation of hotter UV/optical emission at early times after the explosion while the photosphere is still compact, MUSSES2020J exhibited substantially {\it redder} colors ($g - r \gtrsim 0$) four days before optical peak, when the transient luminosity was $\lesssim 10^{44}$ erg s$^{-1}$, than near the peak itself ($g - r \approx -0.5$).  In light of the above discussion, it is tempting to associate the pre-maximum red-to-blue color evolution with the residual absorption by, and subsequent destruction of, pre-existing dusty CSM.  Follow-up photometry and spectra were obtained for this event, but the bluer rest-frame wavelengths they cover $\lambda \lesssim 0.5\mu m$ (due in part to the high source redshift $z \sim 1$) unfortunately preclude constraining the presence of an IR dust echo. 

\acknowledgements

We are grateful to Ryan Chornock, Keiichi Maeda, Raffaella Margutti, Tatsuya Matsumoto for providing detailed comments on the manuscript.  B.D.M.~thanks Kohki Uno for helpful conversations regarding MUSSES2020J which originally motivated this work.  B.D.M.~was supported in part by the National Science Foundation (grant number AST-2009255).  The Flatiron Institute is supported by the Simons Foundation.

\end{document}